\documentclass[12pt]{iopart}

\usepackage{graphicx}
\usepackage{color}
\begin{document}

\title[The floret-pentagonal-lattice Heisenberg antiferromagnet]{
Magnetization process of the $S=1/2$ Heisenberg antiferromagnet
on the floret pentagonal lattice}

\author{Rito~Furuchi$^{1}$, Hiroki~Nakano$^{1}$, Norikazu~Todoroki$^{2}$ and Toru~Sakai$^{1,3}$}

\address{$^{1}$Graduate School of Material Science,
University of Hyogo,
Kamigori,
Hyogo 678-1297, Japan \\
$^{2}$Chiba Institute of Technology, Narashino, Chiba 275-0023, Japan
\\
$^{3}$National Institutes for Quantum and Radiological Science and Technology,
SPring-8
Sayo, Hyogo 679-5148, Japan
\\
}
\ead{dribloodwolf@yahoo.co.jp}
\vspace{10pt}
\begin{indented}
\item[]June 2021
\end{indented}

\begin{abstract}
We study the $S$=1/2 Heisenberg antiferromagnet
on the floret pentagonal lattice by numerical diagonalization method.
This system shows
various behaviours that are different
from that of the Cairo-pentagonal-lattice antiferromagnet.
The ground-state energy without magnetic field
and the magnetization process of this system are reported.
Magnetization plateaux appear
at one-ninth height of the saturation magnetization,
at one-third height, and at seven-ninth height.
The magnetization plateaux at one-third and seven-ninth heights
come from interactions linking the sixfold-coordinated spin sites.
A magnetization jump appears from the plateau at one-ninth height
to the plateau at one-third height.
Another magnetization jump is observed between the heights
corresponding to the one-third and seven-ninth plateaux;
however the jump is away from the two plateaux, namely,
the jump is not accompanied with any magnetization plateaux.
The jump is a peculiar phenomenon that has not been reported.
\end{abstract}

%
%
%
%
%

\section{Introduction}

Frustration is a source of various exotic phenomena in magnetic materials.
One example is a magnetization plateau observed in the magnetization process.
Such frustration in magnetic materials occurs when the systems
have a structure including an odd-number polygon formed
by antiferromagnetic-interaction bonds.
Among such systems, the triangular-lattice system is
extensively studied\cite{review_triangular_lattice_AF} as the most typical case.
The triangular lattice is formed only by congruent regular triangles
of the single type.
The kagome-lattice system is another case that is
widely studied\cite{review_kagome_lattice_AF}
although the kagome lattice includes not only triangles but also hexagons.
The orthogonal dimer system\cite{Shastry_Sutherland_Physica1981,
Kageyama_PRL1999,HN_TSakai_OrthDim1_JPSJ2018}
also includes local triangles and squares.
In such cases including the local triangular structure,
frustration created by the structure plays essential roles
in the behaviour of total magnetic systems.

Among such local structure of an odd-number polygon,
the pentagonal structure is the next candidate.
Unfortunately,
number of studies concerning systems that are composed
of a local pentagonal structure is lower than
those for the triangular structure.
The Cairo-pentagonal-lattice system is
a precious example\cite{Ressouche_PRL2009,
Rousochatzakis_Cairo,HNakano_Cairo_lt,
Isoda_Cairo_full}.
The Cairo pentagonal lattice is formed only by congruent pentagons
of the single type without remaining spaces.
If we consider the tiling problem within the single plane,
all the five edges of a congruent pentagon
in the Cairo pentagonal lattice can be of equivalent length
although the pentagon is not a regular one.
The Cairo-pentagonal-lattice Heisenberg antiferromagnet
reveals magnetization plateau at one-third of the saturation magnetization
in its magnetization process. The plateau is accompanied by
a characteristic magnetization jump in specific parameter regions.
Not only theoretical studies but also experimental reports
have been reported\cite{Abakumov_PRB2013,Tsirlin_PRB2017,
Chattopadhyay_SciRep2017,Cumby_DaltonTrans2016,Beauvois_PRL2020}
for candidate materials of the Cairo-pentagonal-lattice system
such as Bi$_{2}$Fe$_{4}$O$_{9}$, Bi$_{4}$Fe$_{5}$O$_{13}$F,
and DyMn$_{2}$O$_{5}$.
The $S=1/2$ Ising models on the Cairo-pentagonal lattice were
studied.\cite{Waldor_PL1984,Waldor_ZP1985,Rojas_PRE2012}
As systems including local pentagonal structure,
spherical kagome cluster\cite{Fukumoto_spherical_kagome_PTEP2014},
dodecahedral cluster\cite{Konstantinidis_dodeca_PRB2005,
Konstantinidis_dodeca_JPCM2016} and
icosidodecahedron cluster
\cite{Exler_icosidodecahedron_PRB2003,Schroder_icosidodecahedron_PRB2005}
are also
studied theoretically
although the spatial dimensionality of these systems is not two.
As one-dimensional systems,
the $S=1/2$ Ising-Heisenberg pentagonal chain\cite{Karlova_PRB2018}
was investigated and
the correlated electron systems with a pentagonal geometry
\cite{Gulacsi_IJPB2013,Gulacsi_PML2014,Kovacs_PML2015}
were also examined.
An experimental study was also reported for a material
including local pentagonal structure
of magnetic bonds\cite{Yamaguchi_SR2015}.

\begin{figure*}
\begin{center}
\includegraphics{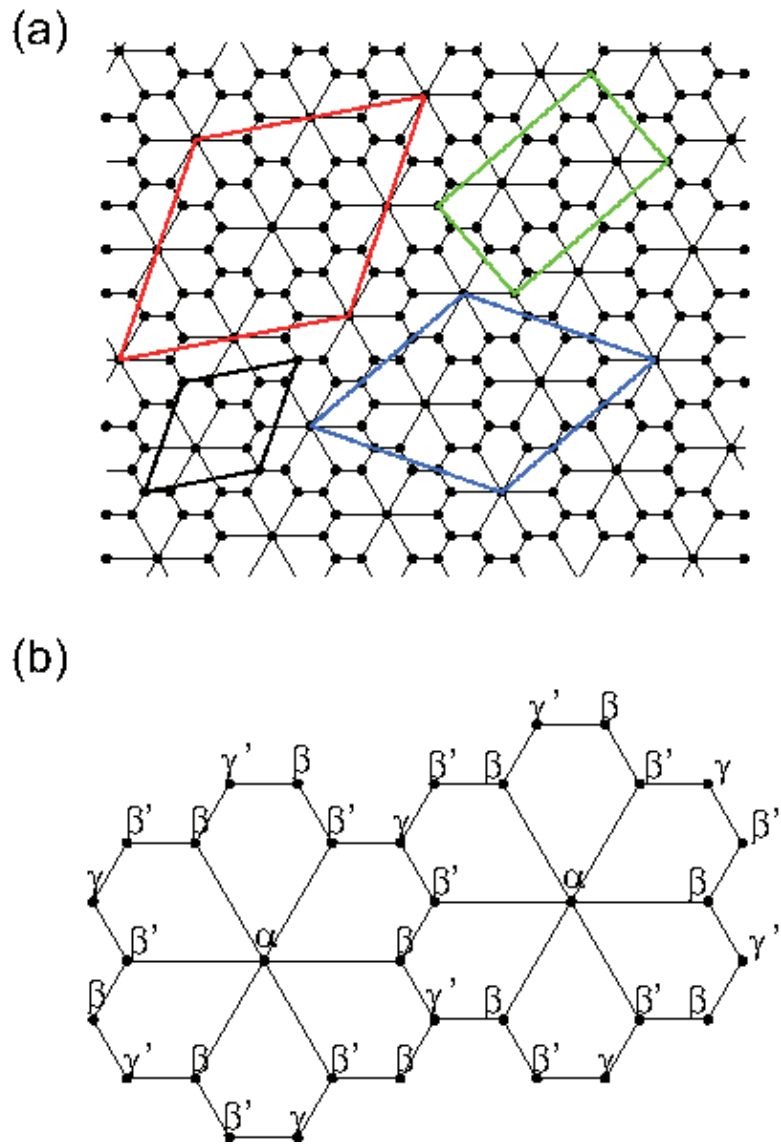}
\end{center}
\caption{ The floret pentagonal lattice is illustrated.
Panel (a) shows its structure and finite-size clusters
treated in this study.
Black, green, blue, and red
dotted lines correspond to the cases for $N=9$, 18, 27, and 36, respectively.
Panel (b) shows neighbouring two unit cells and explains
the correspondence between spin sites and their groups:
$\alpha$, $\beta$, $\beta^{\prime}$, $\gamma$, and $\gamma^{\prime}$.
}
\label{fig1}
\end{figure*}

Under circumstances,
the purpose of this letter is to present the second example
of an investigation of $S=1/2$ Heisenberg antiferromagnet
on the lattice formed only by congruent pentagons.
The lattice is the floret pentagonal lattice,
which is illustrated in Fig.~\ref{fig1}(a).
This lattice is also formed only by congruent pentagons
of the single type without remaining spaces.
When the pentagons are arranged within the single plane
as a tiling problem, the pentagons are not regular ones;
additionally, the lengths of the pentagonal edges are not equivalent.
This lattice has been known in the tiling
problem\cite{Schattschneider1978};
however, the magnetism has not been investigated for a physical system
on this lattice.
The present letter clarifies
characteristic behaviours in the magnetization process
of the antiferromagnet on this lattice.

This paper is organized as follows.
The next section is devoted
to the introduction of the model Hamiltonian and
explanation of our numerical method.
In the third section,
our results will be presented and discussed.
We will first observe the magnetization curve of the classical case.
Next, we will study the quantum case.
The magnetization process will be examined.
The local magnetization will also be discussed.
We will summarize our results and give some remarks in the final section.

\section{Model Hamiltonian and Method of Calculations}

Before introducing the model Hamiltonian,
let us summarize the geometric characteristics
of the floret pentagonal lattice in comparison
with the Cairo pentagonal lattice.
A unit cell of the floret pentagonal lattice includes
nine vertices, which are compared with six vertices
in the case of the Cairo pentagonal lattice.
Vertices of pentagons in the floret pentagonal lattice are
divided into two types:
one is a vertex of the type characterized by the coordination number $z=6$
and
the other is a vertex with $z=3$.
Note here that
the situation of these coordination numbers in the floret pentagonal lattice
is different from the case of the Cairo pentagonal lattice that
reveals $z=3$ and $z=4$.
The situation in the floret pentagonal lattice
is illustrated in Fig.~\ref{fig1}(b).
The vertices of the former type characterized by $z=6$
is called $\alpha$ sites, hereafter.
In addition,
vertices of the latter type of $z=3$ are divided into two groups.
One is a vertex linked by a bond with an $\alpha$ vertex
and the other
is a vertex that is not.
The former (latter) group is composed of vertices
connected by an $\alpha$ vertex
which are called $\beta$ ($\gamma$) sites, hereafter.
Note here that there are two neighbouring $\beta$ sites.
The two spins on these sites favorably reveal different directions
from each other due to the antiferromagnetic interaction between them;
in the view point of spin configurations, $\beta$ sites are
further divided into two groups: $\beta$ and $\beta^{\prime}$.
The division influences $\gamma$ sites which are also
divided into two groups: $\gamma$ and $\gamma^{\prime}$.
We can therefore arrange $\beta$, $\beta^{\prime}$, $\gamma$ and
$\gamma^{\prime}$ on the circumference of the single floret
centering an $\alpha$ site
so that
symbols without prime ($\beta$ and $\gamma$)
and those with prime ($\beta^{\prime}$ and $\gamma^{\prime}$) appear
alternatively as shown in Fig.~\ref{fig1}(b).
Note here that the number of spin sites in a unit cell of this lattice
is 1, 3, 3, 1, and 1 for $\alpha$, $\beta$, $\beta^{\prime}$,
$\gamma$, and $\gamma^{\prime}$, respectively.
The situation of site groups suggests that
a spin configuration realized in the floret-pentagonal-lattice case
is essentially different from that in the Cairo-pentagonal-lattice one.

The Hamiltonian studied in this research is given
by ${\cal H} = {\cal H}_{0} + {\cal H}_{\mathrm{zeeman}}$,
where
\begin{equation}
  {\cal H}_{0}=\sum_{\langle i,j \rangle}
  J \mbox{\boldmath $S$}_{i} \cdot \mbox{\boldmath $S$}_{j}, \
  {\cal H}_{\rm zeeman} = - h \sum_{j} S_{j}^{z} .
\end{equation}
Here, $\mbox{\boldmath $S$}_{i}$ represents the $S = 1/2$ spin operator
at site $i$ illustrated by the closed circle
at a vertex shown in Fig.\ref{fig1}.
The sum with $\langle i,j \rangle$ in ${\cal H}_{0}$
runs over all the pairs of spin sites
linked by the solid lines in Fig.~\ref{fig1}.
Energies are measured in units of $J$.
We examine mainly the case of antiferromagnetic interaction;
so we set $J=1$ hereafter.
The number of spin sites is represented by $N$.
The periodic boundary condition is imposed
for clusters with site $N$, which are shown in Fig.~\ref{fig1}(a).
Note here that the clusters for $N=9$, 27 and 36 are rhombic,
which is related to the three-fold rotational symmetry of the system.
On the other hand, the $N=18$ cluster is not rhombic.
The nonrhombic nature means this cluster has
only a lower rotational symmetry.
In this study, the lowest energy of ${\cal H}_{0}$
is calculated in the subspace characterized by $M$
defined by $\sum_{j} S_{j}^{z}$.
The calculations are carried out based on the Lanczos algorithm
and/or the householder algorithm.
The saturation value of $M$ is given by $M_{\rm s}(=SN)$,
until which $M$ increases discretely with
$\delta M=$1.
The energy is represented by $E(N,M)$.
Part of Lanczos diagonalizations has been carried out
using the MPI-parallelized code, which originally developed
in the research of the Haldane gaps\cite{HNakano_HaldaneGap_JPSJ2009}.
The usefulness of our program
was confirmed in several large-scale parallelized
calculations\cite{HNakano_kgm_gap_JPSJ2011,HNakano_s1tri_LRO_JPSJ2013,
HN_TSakai_kgm_1_3_JPSJ2014,HN_TSakai_kgm_S_JPSJ2015,
HN_YHasegawa_TSakai_dist_shuriken_JPSJ2015,HN_TSakai_dist_tri_JPSJ2017,
HN_TSakai_tri_NN_JPSJ2017,HN_TSakai_kgm45_JPSJ2018,
YHasegawa_HN_TSakai_dist_shuriken_PRB2018,TSakai_HN_ICM018,
HN_TSakai_S2HaldaneGap_JPSJ2018,HN_NTodoroki_TSakai_S5S6HaldaneGap_JPSJ2019}.

\begin{figure*}
\begin{center}
\includegraphics{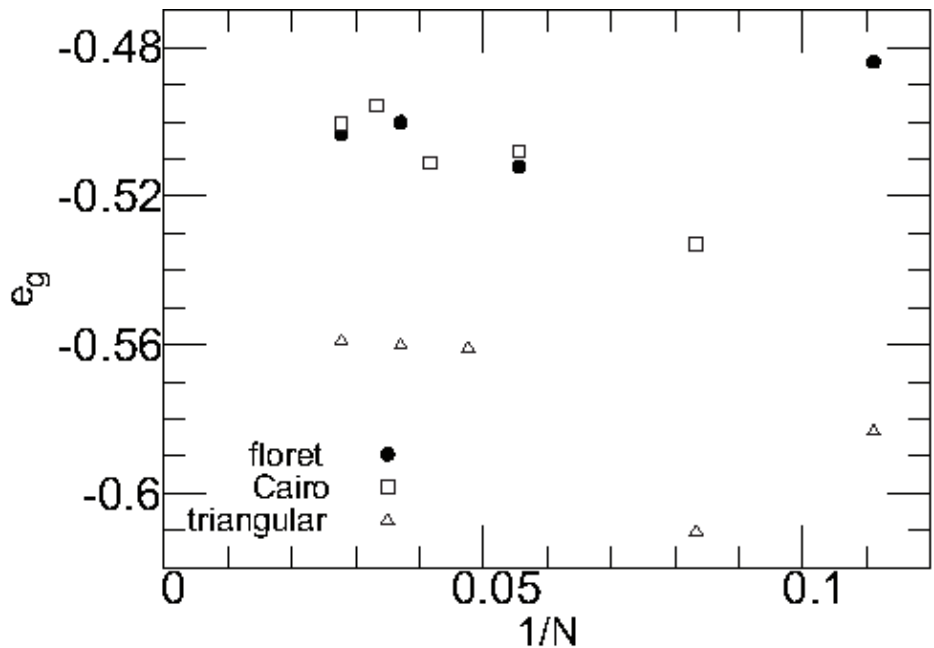}
\end{center}
\caption{The system-size dependence
of the ground-state energy per spin site
for the $S=1/2$ Heisenberg antiferromagnet without magnetic field
as a function of $1/N$.
Closed circles denote results the cases on the floret pentagonal lattice.
For comparison, open squares and open triangles
denote results for the Cairo-pentagonal and triangular lattices.}
\label{fig2}
\end{figure*}

\section{Results and Discussions}

First, we examine the ground-state energy
when the external field is absent.
Figure~\ref{fig2} presents
the ground-state energy per spin site
of the floret-pentagonal-lattice system
in comparison with those
for the Cairo-pentagonal-lattice and triangular-lattice
systems\cite{comment_floret_cairo_triangular}.
Results of $e_{\rm g}$ given by $E(N,0)$ for even $N$
and $E(N,1/2)$ for odd $N$
are shown in Fig.~\ref{fig2} as a function of $1/N$.
Although the data sequence for each system does not show
a monotonic dependence as a consequence of complex finite-size effects,
the Cairo-pentagonal-lattice and triangular systems briefly show
an increasing tendency of the ground-state energy
as $N$ is increased;
on the other hand, the floret-pentagonal-lattice system
does not show an increasing behaviour as $N$ is increased.
The present results suggest that
the appearance of frustration effect in the finite-size clusters of
the floret-pentagonal-lattice system
is different from the other cases.

\begin{figure*}
\begin{center}
\includegraphics{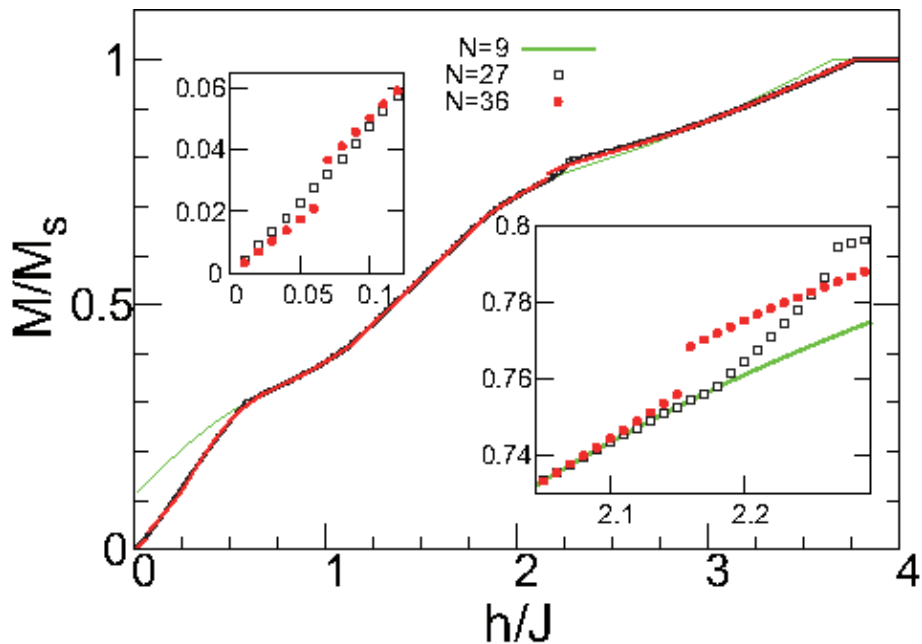}
\end{center}
  \caption{ Magnetization process of the classical case
    of the Heisenberg antiferromagnet
    on the floret pentagonal lattice.
Green solid curve denotes results for $N=9$;
black open squares and red closed circles denote results for $N=27$ and 36,
respectively. 
Insets are zoom-in views in two places, using the same curve and symbols.
  }
\label{fig3}
\end{figure*}

Next, we consider the magnetization process;
before observing the result of the $S=1/2$ model
on the floret pentagonal lattice,
let us examine the magnetization process for the classical model
on the same lattice.
We carry out calculations based on an iterative method
for three rhombic cases illustrated in Fig.\ref{fig1}(a);
results are depicted in Fig.\ref{fig3}.
Note here that
in an intermediate range of $h$,
all the results from the three size samples agree with each other;
results for $N=27$ agree with ones for $N=9$
between $h/J\sim 0.58$ and 2.17, on the other hand,
results for $N=36$ agree with ones for $N=9$
between $h/J\sim 0.68$ and 2.05.
This agreement suggests that classical spin states in this range of $h$
can be understood within the behaviour of the system
of a unit cell of the lattice.
In this intermediate range of $h$,
no significant discontinuous behaviour are observed
in the main panel of Fig.\ref{fig3}
although there also appear two weak kink-like behaviours.
The disagreement outside this intermediate range of $h$
suggests that a unit cell of the spin states for larger systems
is bigger than a unit cell of the lattice.
In the range of fields lower than the intermediate range,
magnetization shows an almost linear behaviour.
No jumps appear for $N=27$;
for $N=36$, however, a jump appears at $h/J \sim 0.07$.
In the range of fields higher than the intermediate range,
a jump around $M/M_{\rm s}\sim 0.79$ appears at $h/J \sim 2.27$ for $N=27$
and
a jump around $M/M_{\rm s}\sim 0.76$ appears at $h/J \sim 2.16$ for $N=36$.
The heights of this jump are close to $M/M_{\rm s}=7/9$,
which will be discussed later.

\begin{figure*}
\begin{center}
\includegraphics{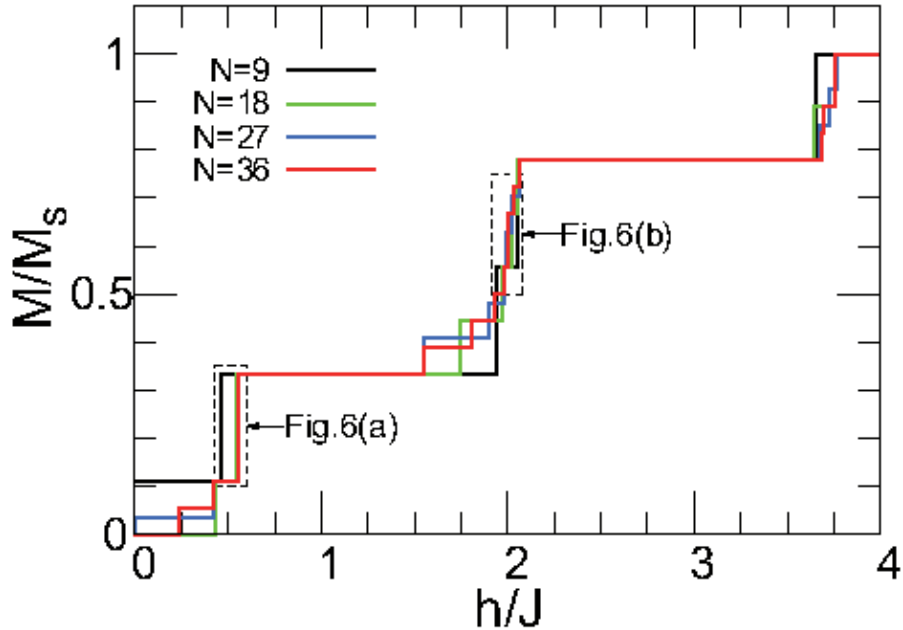}
\end{center}
\caption{
Magnetization process of the $S=1/2$ Heisenberg antiferromagnet
on the floret pentagonal lattice.
Results
after the Maxwell construction
for the entire range from $M=0$ to the saturation magnetization are shown.
Results for $N=9$, 18, 27, and 36
are denoted by
black,
green, blue, and red lines, respectively.
Regarding the two parts surrounded by the broken rectangles,
zoom-in views will be presented in Fig.~\ref{fig6}
to observe well the behaviour of the magnetization jumps
before and after the Maxwell construction.
}
\label{fig4}
\end{figure*}

Now, let us observe the magnetization process
of the $S=1/2$ model on the floret pentagonal lattice.
For a finite-size system, the magnetization process
is determined by the magnetization increase from $M$ to $M+1$
at the field
\begin{equation}
h = E(N,M+1) -  E(N,M),
\label{field_determining}
\end{equation}
under the condition that the lowest-energy state with $M$ and that with $M+1$
become the ground state in specific values of $h$.
If the lowest-energy state with the magnetization $M$ does not become
the ground state in any field,
we should use Maxwell construction to determine
the magnetization process around the magnetization $M$.
The result is shown in Fig.~\ref{fig4}.
The magnetization process exhibits
magnetization plateaux and magnetization jumps.
The next paragraph is devoted to the discussion of the
magnetization plateaux.
After that, we will discuss the magnetization jumps.

\begin{figure*}
\begin{center}
\includegraphics{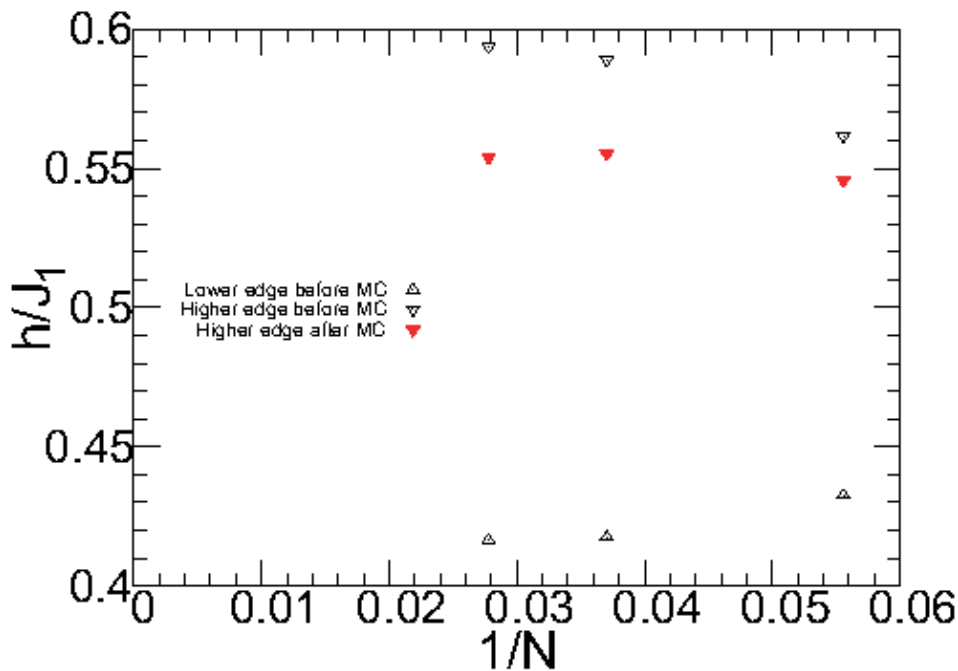}
\end{center}
\caption{System-size dependence of the edges at $M/M_{\rm s}=1/9$
in the finite-size magnetization processes.
Black symbols denote results before the Maxwell construction
(MC);
triangles and inversed triangles represent results
for the lower-field and higher-field edges of this height, respectively.
Red inversed triangles denote results for the higher-field edge of this height
after the Maxwell construction.
}
\label{fig5}
\end{figure*}

Significant behaviour in the magnetization process
is the appearance of the magnetization plateaux at $M/M_{\rm s}=1/3$ and 7/9.
Although it is considered that these plateaux are related
with the number of spin sites in a unit cell of the present lattice,
it is noticeable that these plateaux do not appear in
Fig.~\ref{fig3} for the classical case.
Concerning around $M/M_{\rm s}=7/9$, particularly,
the behaviour of the classical case is a jump
which is completely different from a plateau.
At $M/M_{\rm s}=1/9$,
a plateau behaviour seems to appear
although the width is much smaller than
the widths of the plateaux at $M/M_{\rm s}=1/3$ and 7/9.
In order to observe the behaviour of the size dependence around $M/M_{\rm s}=1/9$
in details, we present Fig.~\ref{fig5} showing the edges at this height
before and after the Maxwell construction.
At the lower-field edge, the Maxwell construction is not necessary;
the field determined by Eq.~\ref{field_determining} show
its size dependence so that the field gets smaller as $N$ is increased.
At the higher-field edge, on the other hand, the Maxwell construction
is carried out; the field after the Maxwell construction shows only
very small size dependence.
Both the behaviours at this height strongly suggest that
the plateau certainly survives when $N$ is infinitely large
in spite of the fact that its width is very narrow.
The magnetization process begins with a state with $M=0$ and
states with $M/M_{\rm s} < 1/9$ are certainly realized;
these states are consistent with the linear behaviour of the classical case.

Another significant behaviour is the appearance of the magnetization jump
between states from $M/M_{\rm s}=1/9$ to $M/M_{\rm s}=1/3$.
For the case of $N=36$,
the behaviour of a magnetization jump also appears at around $M=11$,
that corresponds to $M/M_{\rm s}=11/18$. 
In this paragraph and the next paragraph,
we focus our attention on the former jump.
The latter jump will be discussed after the next paragraph. 
The magnetization jumps have been reported
in the magnetization process of the Heisenberg antiferromagnets
on various two-dimensional lattices:
the kagome-lattice\cite{HN_TSakai_kgm_1_3_JPSJ2014,
HN_TSakai_kgm45_JPSJ2018,Hida_kagome},
square-kagome-lattice\cite{HN_YHasegawa_TSakai_dist_shuriken_JPSJ2015,
YHasegawa_HN_TSakai_dist_shuriken_PRB2018,HN_TSakai_Spin_Flop},
Cairo-pentagonal-lattice\cite{HNakano_Cairo_lt,Isoda_Cairo_full} cases.
The kagome-lattice antiferromagnet shows the jump
at the higher-field-side edge of the $M/M_{\rm s}=5/9$ plateau.
When a specific distortion is switched on in the kagome lattice,
the jump appears at the higher-field-side edge
of the $M/M_{\rm s}=1/3$ plateau.
In the square-kagome-lattice antiferromagnet
that is formed by interaction bonds on the circumference of squares
and interaction bonds on the circumference of octagons,
the jump appears at the higher-field-side edge
of the $M/M_{\rm s}=1/3$ plateau;
it was also reported that the height of this jump
for the square-kagome-lattice antiferromagnet
gets larger as the ratio of
the
two interaction parameters is varied.
The Cairo-pentagonal-lattice antiferromagnet shows a jump
either at the higher-field-side edge
of the $M/M_{\rm s}=1/3$ plateau
or at the lower-field-side edge of the $M/M_{\rm s}=1/3$ plateau;
the side depends on the ratio of two interaction parameters.
In the Cairo-pentagonal-lattice case,
the increasing height of the jump at the lower-field-side edge
of the $M/M_{\rm s}=1/3$ plateau is known;
the jump occurs from $M/M_{\rm s}=1/9$ to $M/M_{\rm s}=1/3$
for a specific parameter case.

\begin{figure*}
\begin{center}
\includegraphics{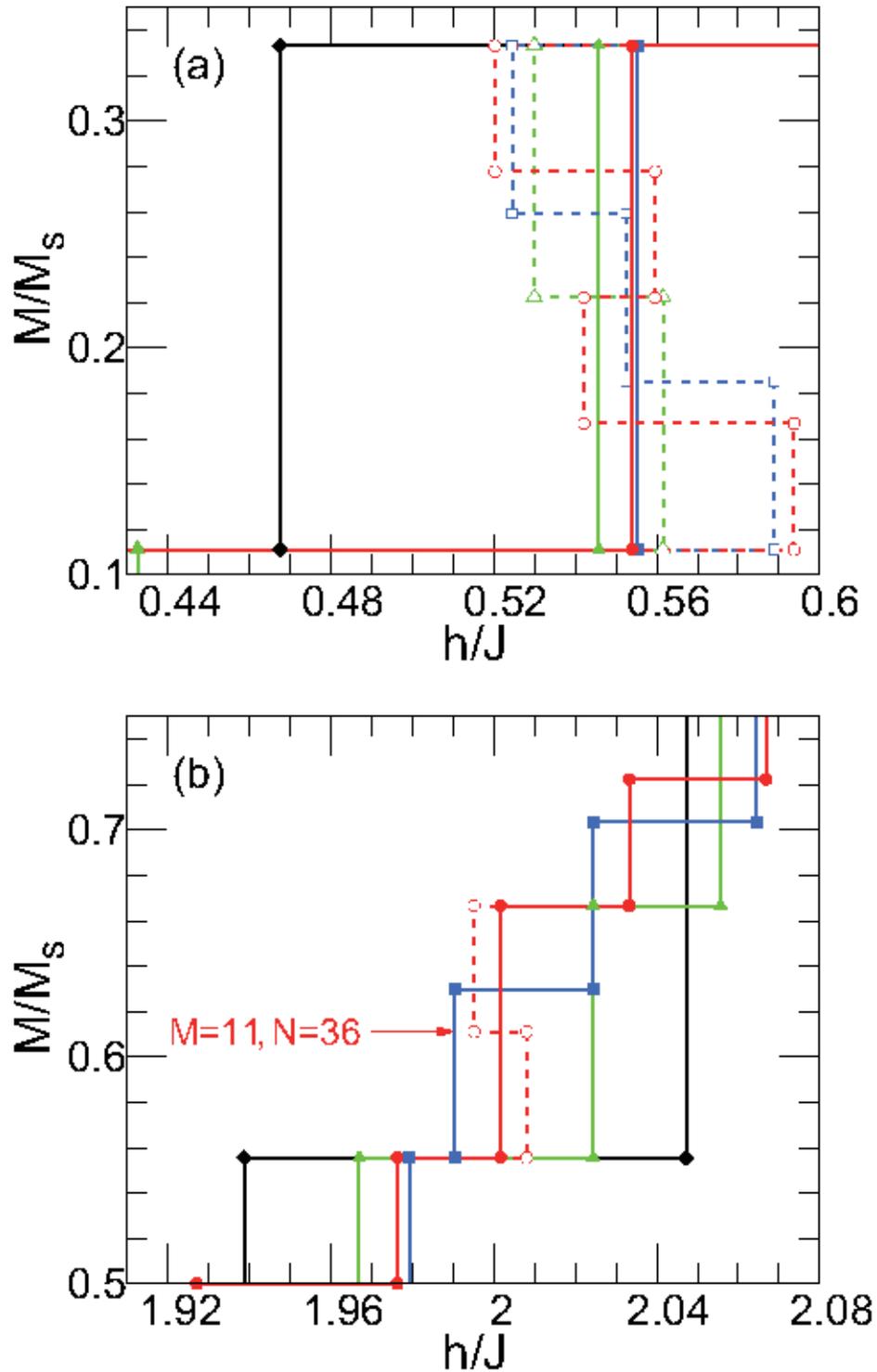}
\end{center}
  \caption{Zoom-in views of Magnetization process
    of the $S=1/2$ Heisenberg antiferromagnet
    on the floret pentagonal lattice.
    Panels (a) and (b) are for the ranges where
    the magnetization jumps appear in Fig.~\ref{fig4}.
The colours are the same as in Fig.~\ref{fig4}.
The solid lines with closed symbols in (a) and (b)
represent the results after the Maxwell construction is carried out.
Black diamonds, blue squares, green triangles, and red circles
in (a) and (b)
correspond to $N=9$, 18, 27, and 36, respectively.
The dotted lines with open symbols in (a) and (b)
represent the results before the Maxwell construction is carried out.
The arrow in (b) indicates the height for $M=11$ and $N_{\rm s}=36$
corresponding to $M/M_{\rm s}=11/18$. 
  }
\label{fig6}
\end{figure*}

Note here that in any previous known cases,
jumps appear at neighbouring location of some magnetization plateaux.
In general, the appearance of the magnetization plateau
at a particular height of $M$ is related to the formation
of a specific quantum state when interactions of the system is varied.
It is because the energy of the state is markedly lowered
due to the changing interactions.
The states in the neighboring heights from $M$ are also influenced
by the same changing interactions; however, degrees of the influence
depend on characteristics of the states outside the plateau.
The change of the energy can be significant although it is not so large
as the plateau state; then, no jumps appear at the edges of the plateau.
On the other hand, there is a possible case
when the change of the energy is quite small.
In this case, a jump appears between the plateau state and
the almost unchanging-energy states outside the plateau.
The jump becomes larger with further change of interactions.
If two plateaux exist and if a jump appears at the edge of each plateau
of the side between the two plateaux,
the two jumps can merge as a consequence of their growth; 
then, the merged jumps becomes a single jump directly from one 
of the two plateaux to the other plateau. 
In this point of view,
the single jump from the plateau at $M/M_{\rm s}=1/9$ 
to the plateau at $M/M_{\rm s}=1/3$ in Fig.~\ref{fig6}(a)
of the present system 
shares the same characteristic 
with the above mentioned Cairo-pentagonal-lattice case.

On the other hand,
the magnetization jump at $M=11$ for $N=36$ in Fig.~\ref{fig6}(b)
of the present system is different from the previously known jumps.
It is a marked characteristic that this jump is not accompanied
with the occurrence of a magnetization plateau.
A jump obeying the mechanism explained in the previous paragraph
is supposed to appear at an edge of an existing plateau.
Although $M/M_{\rm s}=5/9$ is a possible height for the appearance of a plateau
because a unit cell of the present system includes nine spins,
no plateau-like behaviour seems to be observed at $M/M_{\rm s}=5/9$ at all.
In this point of view, the jump in Fig.~\ref{fig6}(b) is peculiar.
Next, let us examine the relationship between the quantum and classical cases
as a possible mechanism of the appearance of the jump in Fig.~\ref{fig6}(b).
In the classical case, there appears a jump around $M/M_{\rm s}=7/9$
in the inset of Fig.~\ref{fig3}.
The height for this classical jump and the one for the quantum jump
at $M=11$ for $N=36$ in Fig.~\ref{fig6}(b) do not correspond to each other.
This disagreement suggests no clear relationship
between the quantum and classical cases.
Concerning the jump in Fig.~\ref{fig6}(b),
it is a priority matter to be resolved
whether or not the jump behaviour at this height survives
as a phenomenon in the thermodynamic limit.
It is difficult to confirm such a macroscopic behaviour
only by numerical-diagonalization studies;
this issue should be tackled by different approaches.
If this jump behaviour is a macroscopic phenomenon,
the mechanism of the appearance should also be investigated intensively.

\begin{figure*}
\begin{center}
\includegraphics{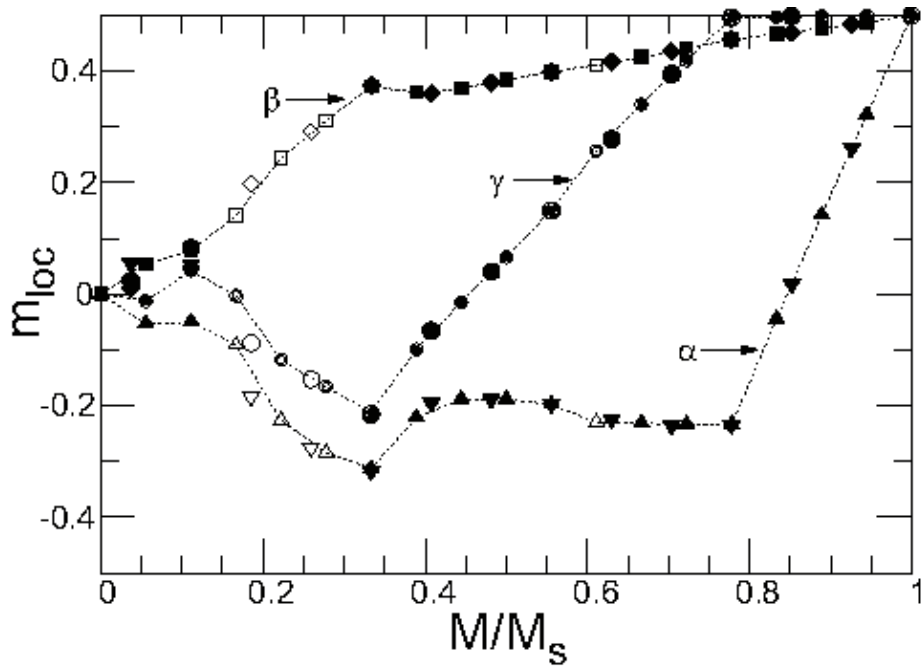}
\end{center}
\caption{Local magnetization of the $S=1/2$ Heisenberg antiferromagnet
on the floret pentagonal lattice under the magnetization field.
Triangles, squares, and double circles linked by dotted lines
denote results of $\alpha$, $\beta$, and $\gamma$ for $N=36$,
respectively.
Inversed triangles, diamonds, and single circles without linking lines
denote results of $\alpha$, $\beta$, and $\gamma$ for $N=27$,
respectively.
Closed symbols denote data for the stably realized states, while open
symbols represent data for the unstable states at the magnetization jump.
}
\label{fig7}
\end{figure*}

To understand the behaviour of the magnetization process
of the $S=1/2$ model on the floret pentagonal lattice,
let us examine how the averaged local magnetization behaves
as the total magnetization is increased.
The averaged local magnetization is evaluated by
\begin{equation}
  m_{\rm loc} = \frac{1}{N_{\xi}}
  \sum_{j\in\xi} \langle S_{j}^{z}\rangle,
\end{equation}
where $\xi$ takes $\alpha$, $\beta$, $\beta^{\prime}$,
$\gamma$, and $\gamma^{\prime}$.
Here the symbol $\langle{\cal O}\rangle$ represents
the expectation value of an operator ${\cal O}$
with respect to the lowest-energy state
within the subspace with a fixed $M$ of interest.
Note here that the average over $\xi$ is carried out
in the case of degenerate ground state for some values of $M$,
where $N_{\xi}$ represents the number of sites belonging to $\xi$.
For cases with the nondegenerate ground state,
results do not change irrespective of whether or not
the averaging is carried out.
Results
for $N=27$ and 36 
are depicted in Fig.~\ref{fig7}.
To avoid a complicated situation, here,
we do not treat the $N=18$ case that does not hold
the three-fold rotational symmetry of the system.
Note here that $m_{\rm loc}$ for $\beta$ and that for $\beta^{\prime}$
agree with each other within their numerical errors;
$m_{\rm loc}$ for $\gamma$ and that for $\gamma^{\prime}$ also agree.
Thus, only results for $\alpha$, $\beta$, and $\gamma$ are shown here.
One can find that system-size dependence of the results in Fig.~\ref{fig7}
is small; therefore,
these results are good for making us understand spin states
of the present system in a field.
When the field is switched on and increased,
our results suggest that
a $\beta$ spin
begins to turn to
the direction of the field
up to $M/M_{\rm s}=1/3$,
that an $\alpha$ spin
begins to turn to
the antiparallel direction of the field,
and that
a $\gamma$ spin initially shows weak dependence of $m_{\rm loc}$ and
turns to the antiparallel direction of the field.
Above $M/M_{\rm s}=1/3$ up to 7/9,
a $\gamma$ spin
begins to turn to
the direction of the field
while $\alpha$ and $\beta$ spins almost stay downward and upward,
respectively.
Above $M/M_{\rm s}=7/9$, finally,
an $\alpha$ spin
begins to turn to
the direction of the field.
The spin states at the magnetization plateaux correspond
to the cases where the change
in the $M$-dependence of $m_{\rm loc}$ appears.
Note here that there is no significant anomalous behaviour
in the dependence at $M=11$ for $N=36$
although the jump appears in the magnetization process.

\begin{figure*}
\begin{center}
\includegraphics{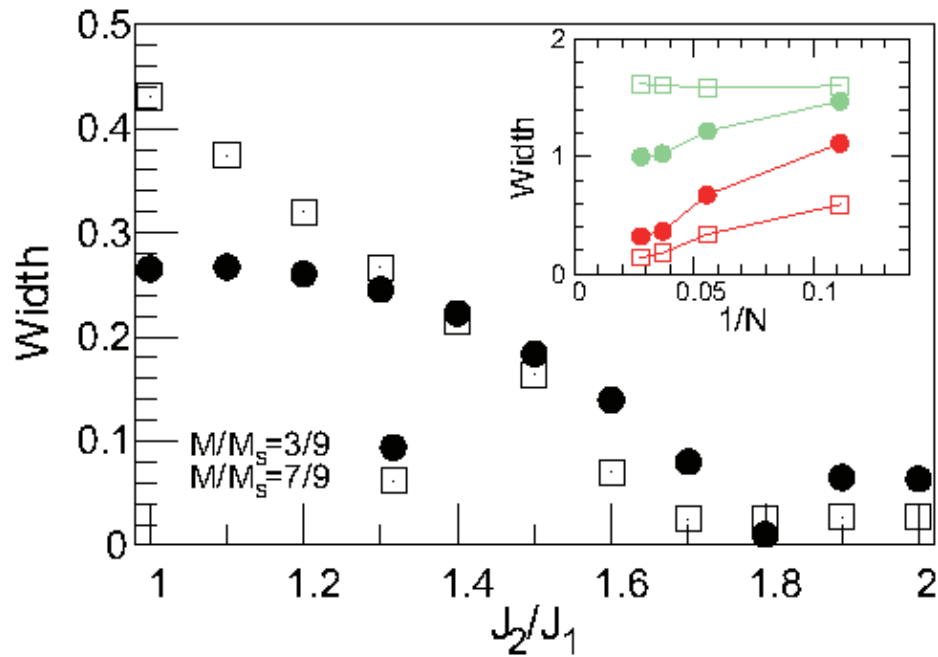}
\end{center}
\caption{
The widths of the specific heights in the magnetization process
of the finite-size systems.
The main panel shows the dependences of the widths
for $N=36$ when the ratio $J_{2}/J_{1}$ is increased from the uniform case;
results at $M/M_{\rm s}=7/9$ and 1/3 are represented
by open squares and closed circles, respectively.
Inset shows the $1/N$-dependences of the widths for
$J_2/J_1$=1 and 2 by green and red symbols, respectively;
symbol types are the same as in the main panel.
}
\label{fig8}
\end{figure*}

In order to examine the possibility of a quantum-origin mechanism
of the plateau that opens in the states at $M/M_{\rm s}=7/9$ and 1/3,
let us consider
the case
when the system includes two kinds of interaction bonds:
one is the bonds between an $\alpha$ site and $\beta$ (or $\beta^{\prime}$) site
and the other is all the rest of the bonds.
The coupling of the former is represented by $J_{1}$ and the latter by $J_{2}$.
  Let us observe the the widths of the heights at $M/M_{\rm s}=7/9$ and 1/3;
results are depicted in Fig.~\ref{fig8}.
The inset of Fig.~\ref{fig8} depicts the system-size dependences
of the widths for $J_2/J_1$=1 and 2.
The dependences for $J_2/J_1$=1 suggest that
the extrapolated values to the thermodynamic limit 
survive
for $M/M_{\rm s}=7/9$ and 1/3.
On the other hand, the results for $J_2/J_1$=2 suggest that
the extrapolated values vanish for both $M/M_{\rm s}$,
which means the absence of the plateaux at these heights of $M/M_{\rm s}$.
The main panel of Fig.~\ref{fig8} depicts
the $J_2/J_1$-dependence of the widths for $N=36$.
Our calculations shows that
when the ratio $J_{2}/J_{1}$ is increased from the uniform case,
the width at $M/M_{\rm s}=7/9$ decreases until $J_{2}/J_{1}\sim 1.6$.
Beyond it, the width maintains a small value,
which agrees with the result for $J_2/J_1$=2 indicating a nonplateau behaviour.
The width at $M/M_{\rm s}=1/3$ also decreases with increasing $J_{2}/J_{1}$
and almost vanishes around $J_{2}/J_{1}\sim 1.8$.
Beyond it, the almost constant narrow width appears
which also suggests to a nonplateau behaviour.
The disappearance of the plateaux suggests that
the plateaux originate from the bonds of $J_{1}$, not the bonds of $J_{2}$.
A possible scenario is the formation of
a local quantum state leading an opening of an energy gap
around an $\alpha$ site.
This possibility should be examined
in future studies.

\section{Summary and Remarks}

We have studied the Heisenberg antiferromagnet
of $S=1/2$ spins on the floret pentagonal lattice in two dimensions.
We have found that its magnetization process shows
the magnetization plateaux at $M/M_{\rm s}=1/9$, 1/3, and 7/9
and that the magnetization jump appears
from the magnetization plateau at $M/M_{\rm s}=1/9$
to the magnetization plateau at $M/M_{\rm s}=1/3$.
We have also found that the appearance of the magnetization jump
at $M=11$ in the $N=36$ system as a peculiar phenomenon
because this jump is not accompanied with any magnetization plateaux.
The mechanism of this jump is still unclear;
this issue should be tackled in future studies.

Note here that, in the Cairo-pentagonal-lattice system,
experimental candidate materials have been reported.
Although the floret pentagonal lattice is even more complicated
from the viewpoint of a structural aspect, it is expected that
candidate materials is experimentally realized.
Further investigations concerning the system on the present lattice
from both theoretical and experimental viewpoints will contribute
much to our understanding frustration effects in magnetic materials.

\ack
We wish to thank Dr. H. Tadano for fruitful discussions.
This work was partly supported by JSPS KAKENHI
Grant Numbers JP16K05418, JP16K05419, JP16H01080 (JPhysics), and
JP18H04330 (JPhysics).
Nonhybrid thread-parallel calculations
in numerical diagonalizations were based on TITPACK version 2
coded by H. Nishimori.
This work used computational resources of
the supercomputer Fugaku
provided by RIKEN
through the HPCI System Research Projects
(Project ID: hp200173, hp210068 and hp210127).
This work used computational resources of
the K computer
provided by RIKEN
through the HPCI System Research Project
(Project ID: hp190053).
This work used computational resources of
Fujitsu PRIMERGY CX600M1/CX1640M1(Oakforest-PACS)
provided by Joint Center for Advanced High Performance Computing
through the HPCI System Research Projects
(Project ID: hp190041 and hp200023).
Some of the computations were performed using facilities
of the Department of Simulation Science,
National Institute for Fusion Science;
Institute for Solid State Physics, The University of Tokyo;
and Supercomputing Division, Information Technology Center,
The University of Tokyo.


\begin{thebibliography}{9}
\bibitem{comment_floret_cairo_triangular}
Between the floret-pentagonal-lattice and
Cairo-pentagonal-lattice systems,
the number of interaction bonds for the same $N$ is common;
note here that
the number of interaction bonds for the triangular-lattice system
is different from the two systems.
\end{thebibliography}
\end{document}